\documentclass[aps,pre,showpacs,amsmath,amssymb]{revtex4-1}
\usepackage{graphicx}

\makeatother

\begin{document}

\title{L\'evy flights in inhomogeneous environments and $1/f$ noise}

\author{R. Kazakevi\v{c}ius}

\email{rytis.kazakevicius@tfai.vu.lt}

\affiliation{Institute of Theoretical Physics and Astronomy, Vilnius University,
A. Go\v{s}tauto 12, LT-01108 Vilnius, Lithuania}

\author{J. Ruseckas}

\affiliation{Institute of Theoretical Physics and Astronomy, Vilnius University,
A. Go\v{s}tauto 12, LT-01108 Vilnius, Lithuania}

\begin{abstract}
Complex dynamical systems which are governed by anomalous diffusion
often can be described by Langevin equations driven by L\'evy stable
noise. In this article we generalize nonlinear stochastic differential
equations driven by Gaussian noise and generating signals with $1/f$
power spectral density by replacing the Gaussian noise with a more
general L\'evy stable noise. The equations with the Gaussian noise
arise as a special case when the index of stability $\alpha=2$. We
expect that this generalization may be useful for describing $1/f$
fluctuations in the systems subjected to L\'evy stable noise.
\end{abstract}

\pacs{05.40.Fb, 05.40.-a, 89.75.Da}

\maketitle

\section{Introduction}

The L\'evy $\alpha$-stable distributions, characterized by the index
of stability $0<\alpha\leqslant2$, constitute the most general class
of stable processes. The Gaussian distribution is their special case,
corresponding to $\alpha=2$. If $\alpha<2$, the L\'evy stable distributions
have power-law tails $\sim1/x^{1+\alpha}$. There are many systems
exhibiting L\'evy $\alpha$-stable distributions: distribution function
of turbulent magnetized plasma emitters \cite{Marandet2003} and step-size
distribution of photons in hot vapors of atoms \cite{Mercadier2009}
have L\'evy tails; theoretical models suggest that velocity distribution
of particles in fractal turbulence is L\'evy stable distribution
\cite{Takayasu1984} or at least has L\'evy tails \cite{Min1996}.
If system behavior depends only on large noise fluctuations, such
noise intensity distributions can by approximated by L\'evy stable
distribution, leading to L\'evy flights. L\'evy flight is a generalization
of the Brownian motion which describes the motion of small macroscopic
particles in a liquid or a gas experiencing unbalanced bombardments
due to surrounding atoms. The Brownian motion mimics the influence
of the ``bath'' of surrounding molecules in terms of time-dependent
stochastic force which is commonly assumed to be white Gaussian noise.
That postulate is compatible with the assumption of a short correlation
time of fluctuations, much shorter than the time scale of the macroscopic
motion, and the assumption of weak interactions with the bath. In
contrast, the L\'evy motions describe results of strong collisions
between the particle and the surrounding environment. L\'evy flights
can be found in many physical systems: as an example we can point
out anomalous diffusion of Na adatoms on solid Cu surface \cite{Guantes2001},
anomalous diffusion of a gold nanocrystal, adsorbed on the basal plane
of graphite \cite{Luedtke1999} and anomalous diffusion in optical
lattices \cite{Marksteiner1996}. L\'evy flights can be modeled by
fractional Fokker-Planck equations \cite{Fogedby1994} or Langevin
equations with additive L\'evy stable noise.

Nonlinear stochastic differential equations (SDEs) with additive L\'evy
stable noise have been explored quite extensive for past 15 years
\cite{Jespersen1999,Chechkin2002,Eliazar2003,Denisov2008a}. Such
stochastic differential equations lead to fractional Fokker-Planck
equations with constant diffusion coefficient. Models with multiplicative
L\'evy stable noise have been used for modeling inhomogeneous media
\cite{Srokowski2009}, ecological population density with fluctuating
volume of resources \cite{Alexander2012}. The relation between Langevin
equation with multiplicative L\'evy stable noise and fractional Fokker-Planck
equation has been introduced in Ref.~\cite{Schertzer2001}, where
Langevin equation is interpreted in It\^o sense \cite{Srokowski2009IS}.
The relation between these two equation are not known in Stratonovich
interpretation. Fractional Fokker-Planck equation models have been
applied to model enzyme diffusion on polymer chain \cite{Lomholt2005}
and some cases of anomalous diffusion \cite{Srokowski2006}. However,
application of L\'evy stable noise driven SDEs can be problematic.
We can always write Fokker-Planck equation corresponding to Langevin
equation driven by Gaussian noise and vice versa, but such statement
is not always true for Langevin equation with L\'evy stable noise.
For example, particle (enzyme) dispersion on rapidly folding random
heteropolymer can be described by space fractional Fokker-Planck equation
\cite{Brockmann2003}, but for such equation counterpart Langevin
equation has not been found \cite{Brockmann2002} and might not even
exits \cite{Brockmann2003lf}.

One of the characteristics of the signal is the power spectral density
(PSD). Signals having the PSD at low frequencies $f$ of the form
$S(f)\sim1/f^{\beta}$ with $\beta$ close to $1$ are commonly referred
to as ``$1/f$ noise'', ``$1/f$ fluctuations'', or ``flicker noise''.
Power-law distributions of spectra of signals with $0.5<\beta<1.5$,
as well as scaling behavior are ubiquitous in physics and in many
other fields \cite{Scholarpedia2007,*Weissman1988,*Barabasi1999,*Gisiger2001,*Wagenmakers2004,*Szabo2007,*Castellano2009}.
Despite the numerous models and theories proposed since its discovery
more than 80 years ago \cite{Johnson1925,*Schottky1926}, the subject
of $1/f$ noise remains still open for new discoveries. Most models
and theories of $1/f$ noise are not universal because of the assumptions
specific to the problem under consideration. A short categorization
of the theories and models of $1/f$ noise is presented in the introduction
of the paper \cite{Kaulakys2009}, see also recent review by Balandin
\cite{Balandin2013}. Mostly $1/f$ noise is considered as Gaussian
process \cite{Kogan2008,Li2012}, but sometimes the signal exhibiting
$1/f$ fluctuations are non-Gaussian \cite{Orlyanchik2008,Melkonyan2010}.

Often $1/f$ noise is modeled as the superposition of Lorentzian spectra
with a wide range distribution of relaxation times \cite{McWhorter1957}.
An influential class of the models of $1/f$ noise involves self-organized
criticality (SOC) \cite{Bak1987,*Jensen1989,*Kertesz1990}. One more
way of obtaining $1/f$ noise from a signal consisting of pulses has
been presented in \cite{Kaulakys1998,*Kaulakys1999,*Kaulakys2000-2,*Kaulakys2005}:
it has been shown that the intrinsic origin of $1/f$ noise may be
a Brownian motion of the interevent time of the signal pulses. The
nonlinear SDEs generating signals with $1/f$ noise were obtained
in Refs.~\cite{Kaulakys2004,Kaulakys2006} (see also papers \cite{Kaulakys2009,Ruseckas2010,*Ruseckas2011}),
starting from the point process model of $1/f$ noise. A special case
of this SDE has been obtained using Kirman's agent model \cite{Ruseckas2011epl}.
Such nonlinear SDEs were used to describe signals in socio-economical
systems \cite{Gontis2010,Mathiesen2013}.

The purpose of this paper is to generalize nonlinear SDEs driven by
Gaussian noise and generating signals with $1/f$ PSD by replacing
the Gaussian noise with a more general L\'evy stable noise. The previously
proposed SDEs then arise as a special case when $\alpha=2$. We can
expect that this generalization may be useful for describing $1/f$
fluctuations in the systems subjected to L\'evy stable noise.

The paper is organized as follows: In Section~\ref{sec:power} we
search for the nonlinear SDE with L\'evy stable noise yielding power
law steady state probability density function (PDF) of the generated
signal. In Section~\ref{sec:spectrum} we estimate when the signal
generated by such an SDE has $1/f$ PSD in a wide region of frequencies.
In Section~\ref{sec:numerical} we numerically solve obtained equations
and compare the PDF and PSD of the signal with analytical estimations.
Section~\ref{sec:concl} summarizes our findings.

\section{Stochastic differential equation with L\'evy stable noise generating
signals with power law distribution}

\label{sec:power}In this Section we search for nonlinear SDEs with
L\'evy stable noise yielding power law steady state PDF of the generated
signal. We consider the Langevin equation of the form \cite{Fogedby1994,Fogedby1994a,Fogedby1998}
\begin{equation}
\frac{dx}{dt}=a(x)+b(x)\xi(t)\,,\label{eq:sde1}
\end{equation}
where $a(x)$ and $b(x)$ are given functions describing the deterministic
drift term and the amplitude of the noise, respectively. The stochastic
force $\xi(t)$ is uncorrelated, $\langle\xi(t)\xi(t')\rangle=\delta(t-t')$
and is characterized by L\'evy $\alpha$-stable distribution. In
this paper we will restrict our investigation only to symmetric stable
distributions, thus the characteristic function of $\xi(t)$ is
\begin{equation}
\langle\exp(ik\xi)\rangle=\exp(-\sigma^{\alpha}|k|^{\alpha})\,.\label{eq:charact}
\end{equation}
Here $\alpha$ is the index of stability and $\sigma$ is the scale
parameter. We interpret Eq.~(\ref{eq:sde1}) in It\^o sense. In
mathematically more formal way Eq.~(\ref{eq:sde1}) can be written
in the form
\begin{equation}
dx=a(x)+b(x)dL_{t}^{\alpha}\,,
\end{equation}
where $dL_{t}^{\alpha}$ stands for the increments of L\'evy $\alpha$-stable
motion $L_{t}^{\alpha}$ \cite{Janicki1994,Weron2005}. For calculating
of the steady state PDF of the signal $x$ we will use the fractional
Fokker-Planck equation instead of stochastic differential equation
(\ref{eq:sde1}). The fractional Fokker-Planck equation corresponding
to It\^o solution of Eq.~(\ref{eq:sde1}) is \cite{Ditlevsen1999,Schertzer2001}
\begin{equation}
\frac{\partial}{\partial t}P(x,t)=-\frac{\partial}{\partial x}a(x)P(x,t)+\sigma^{\alpha}\frac{\partial^{\alpha}}{\partial|x|^{\alpha}}b(x)^{\alpha}P(x,t)\,.\label{eq:fracFP}
\end{equation}
Here $\partial^{\alpha}/\partial|x|^{\alpha}$ is the Riesz-Weyl fractional
derivative. The Riesz-Weyl fractional derivative of the function $f(x)$
is defined by its Fourier transform \cite{Samko1993},
\begin{equation}
\mathcal{F}\left[\frac{\partial^{\alpha}}{\partial|x|^{\alpha}}f(x)\right]=-|k|^{\alpha}\tilde{f}(k)\,.
\end{equation}
One can get the following expression for the Riesz-Weyl derivative
:
\begin{equation}
\frac{\partial^{\alpha}}{\partial|x|^{\alpha}}f(x)=-\frac{1}{2\cos\left(\frac{\pi\alpha}{2}\right)}\{D_{+}^{-\alpha}f(x)+D_{-}^{-\alpha}f(x)\}\,,\label{eq:R-W}
\end{equation}
where $D_{+}^{-\alpha}$ and $D_{-}^{-\alpha}$ are the left and right
Riemann-Liouville derivatives \cite{Samko1993}:
\begin{equation}
D_{\pm}^{-\alpha}=(\pm1)^{m}\frac{d^{m}}{dx^{m}}D_{\pm}^{m-\alpha}\,,\qquad m-1<\alpha<m\,.
\end{equation}
Here $m$ is an integer and
\begin{eqnarray}
D_{+}^{\alpha}f(x) & = & \frac{1}{\Gamma(\alpha)}\int_{-\infty}^{x}(x-z)^{\alpha-1}f(z)\, dz\,,\\
D_{-}^{\alpha}f(x) & = & \frac{1}{\Gamma(\alpha)}\int_{x}^{+\infty}(z-x)^{\alpha-1}f(z)\, dz\,.
\end{eqnarray}
When $\alpha=1$ then the definition of the Riesz-Weyl derivative
is
\begin{equation}
\frac{d}{d|x|}f(x)=-\frac{d}{dx}\frac{1}{\pi}\int_{-\infty}^{+\infty}\frac{f(z)}{x-z}\, dz\,.
\end{equation}
Eq.~(\ref{eq:fracFP}) leads to the following equation for the steady
state PDF:
\begin{equation}
\sigma^{\alpha}\frac{\partial^{\alpha}}{\partial|x|^{\alpha}}b(x)^{\alpha}P_{0}(x)-\frac{\partial}{\partial x}a(x)P_{0}(x)=0\,.\label{eq:stationary}
\end{equation}
Equation (\ref{eq:stationary}) can be written as $-dJ(x)/dx=0$,
where $J(x)$ is the probability current. Reflective boundaries lead
to the boundary condition $J(x)=0$.

\subsection{Equation with only positive values of $x$}

We will search for the stochastic differential equation (\ref{eq:sde1})
generating signals with power law steady state PDF, 
\begin{equation}
P_{0}(x)\sim x^{-\lambda}\,.
\end{equation}
Since power law PDF cannot be normalized when $x$ can vary from zero
to infinity, we will assume that the power law holds only in some
wide region of $x$, $x_{\mathrm{min}}\ll x\ll x_{\mathrm{max}}$.
One can expect that power law PDF can be obtained when the coefficients
$a(x)$ and $b(x)$ in Eq.~(\ref{eq:sde1}) themselves are of the
power law form. Thus we will consider $b(x)=x^{\eta}$ and $a(x)=\sigma^{\alpha}\gamma x^{\mu}$.
Here $\eta$ is the exponent of the multiplicative noise, $\mu$ and
$\gamma$ are to be determined. With such a choice of $b(x)$ and
power law form of $P_{0}(x)$ from Eq.~(\ref{eq:fracFP}) it follows
that we need to calculate fractional derivative of the the power law
function.

Let us consider the function
\begin{equation}
f(x)=\begin{cases}
x^{\rho}\,, & x_{\mathrm{min}}<x<x_{\mathrm{max}}\,,\\
0 & \mbox{otherwise .}
\end{cases}\label{eq:f1}
\end{equation}
Using Eq.~(\ref{eq:R-W}) we obtain the following approximate expressions
for the fractional derivative of the function (\ref{eq:f1}) when
$x_{\mathrm{min}}\ll x\ll x_{\mathrm{max}}$:
\begin{equation}
\frac{d^{\alpha}}{d|x|^{\alpha}}f(x)\approx\begin{cases}
\frac{\sin\left(\pi\left(\frac{\alpha}{2}-\rho\right)\right)}{\sin(\pi(\rho-\alpha))}\frac{\Gamma(1+\rho)}{\Gamma(1+\rho-\alpha)}x^{\rho-\alpha}\,, & -1<\rho<\alpha\,,\\
\frac{x_{\mathrm{min}}^{1+\rho}}{2\cos\left(\frac{\pi}{2}\alpha\right)(1+\rho)\Gamma(-\alpha)}x^{-1-\alpha}\,, & \rho<-1\\
\frac{x_{\mathrm{max}}^{\rho-\alpha}}{2\cos\left(\frac{\pi}{2}\alpha\right)(\alpha-\rho)\Gamma(-\alpha)}\,, & \rho>\alpha
\end{cases}\,,\qquad0<\alpha<2;\,\alpha\neq1\label{eq:Riesz-power}
\end{equation}
and
\begin{equation}
\frac{d}{d|x|}f(x)\approx\begin{cases}
-\lambda\cot(\pi\rho)x^{\rho-1}\,, & -1<\rho<1\\
-\frac{x_{\mathrm{min}}^{1+\rho}}{\pi(1+\rho)}x^{-2}\,, & \rho<-1\\
\frac{x_{\mathrm{max}}^{\rho-1}}{\pi(1-\rho)}\,, & \rho>1
\end{cases}
\end{equation}
for $\alpha=1$. We see that the approximate expression for the fractional
derivative does not depend on the limiting values $x_{\mathrm{min}}$
and $x_{\mathrm{max}}$ when $-1<\rho<\alpha$. Using the power-law
forms of the coefficients $a(x)$ and $b(x)$, assuming that $-1<\alpha\eta-\lambda<\alpha$
and using Eq.~(\ref{eq:Riesz-power}) for the fractional derivative,
from Eq.~(\ref{eq:stationary}) we get
\begin{equation}
\frac{\sin\left[\pi\left(\frac{\alpha}{2}-\alpha\eta+\lambda\right)\right]}{\sin[\pi(\alpha(\eta-1)-\lambda)]}\frac{\Gamma(1+\alpha\eta-\lambda)}{\Gamma(1+\alpha(\eta-1)-\lambda)}x^{\alpha(\eta-1)-\lambda}-\gamma(\mu-\lambda)x^{\mu-\lambda-1}=0\,.
\end{equation}
This equation should be valid for all values of $x$. This can be
only when
\begin{equation}
\mu=\alpha(\eta-1)+1
\end{equation}
and
\begin{equation}
\gamma=\frac{\sin\left[\pi\left(\frac{\alpha}{2}-\alpha\eta+\lambda\right)\right]}{\sin[\pi(\alpha(\eta-1)-\lambda)]}\frac{\Gamma(\alpha\eta-\lambda+1)}{\Gamma(\alpha(\eta-1)-\lambda+2)}\,.\label{eq:gamma-pos}
\end{equation}
Thus we will investigate the nonlinear SDE with L\'evy stable noise
of the form
\begin{equation}
dx=\sigma^{\alpha}\frac{\sin\left[\pi\left(\frac{\alpha}{2}-\alpha\eta+\lambda\right)\right]}{\sin[\pi(\alpha(\eta-1)-\lambda)]}\frac{\Gamma(\alpha\eta-\lambda+1)}{\Gamma(\alpha(\eta-1)-\lambda+2)}x^{\alpha(\eta-1)+1}dt+x^{\eta}dL_{t}^{\alpha}\,.\label{eq:sde-power-1}
\end{equation}
This equation is a generalization of the nonlinear SDE with Gaussian
noise proposed in Refs.~\cite{Kaulakys2004,Kaulakys2006}. Because
of the divergence of the power law distribution and the requirement
of the stationarity of the process, the SDE (\ref{eq:sde-power-1})
should be analyzed together with the appropriate restrictions of the
diffusion in some finite interval. The simplest choice of restriction
is the reflective boundaries at $x=x_{\mathrm{min}}$ and $x=x_{\mathrm{max}}$.
However, other forms of restrictions are possible by introducing additional
terms in the drift term of Eq.~(\ref{eq:sde-power-1}).

From Eq.~(\ref{eq:Riesz-power}) it follows that the equation for
the fractional derivative is valid when $-1<\alpha\eta-\lambda<\alpha$.
However, the condition $J(x)=0$ for the probability current leads
to a stronger restriction than Eq.~(\ref{eq:stationary}) which ensures
only $dJ(x)/dx=0$. Using Eq.~(\ref{eq:R-W}) and the function (\ref{eq:f1})
we see that the upper limiting value $x_{\mathrm{max}}$ can be neglected
in the probability current when $\rho<\alpha-1$. Thus the power law
exponent $\lambda$ of the steady state PDF should be from the interval
\begin{equation}
\alpha(\eta-1)+1<\lambda<\alpha\eta+1\,.\label{eq:cond_pow_law-1}
\end{equation}
As a particular case when $\alpha=2$ from Eq.~(\ref{eq:sde-power-1})
we get previously proposed SDE with the Gaussian noise \cite{Kaulakys2004,Kaulakys2006}
\begin{equation}
dx=\sigma^{2}(2\eta-\lambda)x^{2\eta-1}dt+x^{\eta}dL_{t}^{2}\,.
\end{equation}
Note, that according to the definition (\ref{eq:charact}), the scale
parameter $\sigma$ differs from the standard deviation of the Gaussian
noise. Eq.~(\ref{eq:sde-power-1}) has a simple form when $\alpha=1$:
\begin{equation}
dx=\sigma\cot[\pi(\lambda-\eta)]x^{\eta}dt+x^{\eta}dL_{t}^{1}\,.
\end{equation}

\subsection{Equations allowing both positive and negative values of $x$}

In Eq.~(\ref{eq:sde-power-1}) the stochastic variable $x$ can acquire
only positive values. Similarly as in Ref.~\cite{Ruseckas2011} we
can get the equations allowing $x$ to be negative. We will search
for the stochastic differential equation (\ref{eq:sde1}) generating
signals with power law steady state PDF 
\begin{equation}
P_{0}(x)\sim|x|^{-\lambda}\,.
\end{equation}
To have a normalizable PDF we will assume that the power law holds
only in some wide region of $x$, $x_{\mathrm{min}}\ll|x|\ll x_{\mathrm{max}}$.
In order to obtain such an equation we will consider Eq.~(\ref{eq:sde1})
with the coefficients having the power law form $a(x)=\sigma^{\alpha}\gamma|x|^{\mu-1}x$
and $b(x)=|x|^{\eta}$ when $|x|\gg x_{\mathrm{min}}$. Similarly
as in the case of the positive $x$ we investigate the fractional
derivative of the function
\begin{equation}
f(x)=\begin{cases}
|x|^{\rho}\,, & x_{\mathrm{min}}<|x|<x_{\mathrm{max}}\,,\\
x_{\mathrm{min}}^{\rho}\,, & -x_{\mathrm{min}}<x<x_{\mathrm{min}}\,,\\
0 & \mbox{otherwise.}
\end{cases}\label{eq:f2}
\end{equation}
Using Eq.~(\ref{eq:R-W}) we obtain the following approximate expressions
for the fractional derivative of the function (\ref{eq:f1}) when
$x_{\mathrm{min}}\ll x\ll x_{\mathrm{max}}$: 
\begin{equation}
\frac{d^{\alpha}}{d|x|^{\alpha}}f(x)\approx\frac{\sin\left(\frac{\pi}{2}\rho\right)}{\sin\left(\frac{\pi}{2}(\alpha-\rho)\right)}\frac{\Gamma(1+\rho)}{\Gamma(1+\rho-\alpha)}x^{\rho-\alpha}\,,\qquad-1<\rho<\alpha\,.\label{eq:frac-deriv}
\end{equation}
Using Eq.~(\ref{eq:frac-deriv}) for the fractional derivative in
Eq.~(\ref{eq:stationary}), we obtain $\mu=\alpha(\eta-1)+1$ and
\begin{equation}
\gamma=\frac{\sin\left[\frac{\pi}{2}(\alpha\eta-\lambda)\right]}{\sin\left[\frac{\pi}{2}(\lambda-\alpha(\eta-1))\right]}\frac{\Gamma(\alpha\eta-\lambda+1)}{\Gamma(\alpha(\eta-1)-\lambda+2)}\,.\label{eq:gamma-pos-neg}
\end{equation}
In addition, from Eq.~(\ref{eq:frac-deriv}) it follows that the
power law exponent $\lambda$ of the steady state PDF should be from
the interval
\begin{equation}
\alpha(\eta-1)<\lambda<\alpha\eta+1\,.
\end{equation}

When $\alpha=2$, Eq.~(\ref{eq:gamma-pos-neg}) simplifies to
\begin{equation}
\gamma=2\eta-\lambda\,.
\end{equation}
This expression is the same as the one for the SDE with only positive
values of $x$ and $\alpha=2$. However, when $\alpha<2$, the coefficient
$\gamma$ given by Eq.~(\ref{eq:gamma-pos-neg}) is different from
$\gamma$ given by Eq.~(\ref{eq:gamma-pos}), in contrast to the
Gaussian case ($\alpha=2$). This can be understood by noticing that
the L\'evy stable noise for $\alpha<2$ has large jumps. Jumps from
the regions with negative values of the stochastic variable $x$ to
the regions with positive values influence the PDF $P_{0}(x)$ for
the positive values of $x$. The same situation is with the jumps
from positive to negative regions. Eq.~(\ref{eq:gamma-pos-neg})
also has a simple form
\begin{equation}
\gamma=\tan\left[\frac{\pi}{2}(\eta-\lambda)\right]
\end{equation}
for $\alpha=1$.

The required form of the coefficients $\alpha(x)$ and $b(x)$ has
the equation
\begin{equation}
dx=\sigma^{\alpha}\gamma(x_{0}^{2}+x^{2})^{\frac{\alpha}{2}(\eta-1)}xdt+(x_{0}^{2}+x^{2})^{\frac{\eta}{2}}dL_{t}^{\alpha}\label{eq:sde-3}
\end{equation}
and equation
\begin{equation}
dx=\sigma^{\alpha}\gamma(x_{0}^{\alpha}+|x|^{\alpha})^{\eta-1}xdt+(x_{0}^{\alpha}+|x|^{\alpha})^{\frac{\eta}{\alpha}}dL_{t}^{\alpha}\,.\label{eq:sde-3a}
\end{equation}
Here parameter $x_{0}$ plays the role of $x_{\mathrm{min}}$. The
restriction the diffusion at the large absolute values of $x$ can
be achieved by reflective boundaries at $\pm x_{\mathrm{max}}$ or
by additional terms in the equations. Eq.~(\ref{eq:sde-3}) is a
generalization of SDE with Gaussian noise from Ref.~\cite{Ruseckas2011}.
The addition of the parameter $x_{0}$ restricts the divergence of
the power law distribution of $x$ at $x\rightarrow0$. Eqs.~(\ref{eq:sde-3}),
(\ref{eq:sde-3a}) for $|x|\ll x_{0}$ represents SDEs with additive
L\'evy stable noise and linear relaxation.

\section{Power spectral density of the generated signals }

\label{sec:spectrum}In this Section we estimate the PSD of the signals
generated by the SDE with L\'evy stable noise
\begin{equation}
dx=\sigma^{\alpha}\gamma x^{\alpha(\eta-1)+1}dt+x^{\eta}dL_{t}^{\alpha}\,,\label{eq:sde-2}
\end{equation}
proposed in the previous Section. Here $\gamma$ is given by Eq.~(\ref{eq:gamma-pos}).
For this estimation we use the (approximate) scaling properties of
the signals, as it is done in the Appendix~A of Ref.~\cite{RuseckasChaos2013}
and in Ref.~\cite{Ruseckas2014}. Using Wiener-Khintchine theorem
the PSD can be related to the autocorrelation function $C(t)$, which
can be calculated using the steady state PDF $P_{0}(x)$ and the transition
probability $P(x',t|x,0)$ (the conditional probability that at time
$t$ the signal has value $x'$ with the condition that at time $t=0$
the signal had the value $x$) \cite{Gardiner2004}:
\begin{equation}
C(t)=\int dx\int dx^{\prime}\, xx^{\prime}P_{0}(x)P(x^{\prime},t|x,0)\,.
\end{equation}
The transition probability can be obtained from the solution of the
fractional Fokker-Planck equation (\ref{eq:fracFP}) with the initial
condition $P(x',t=0|x,0)=\delta(x'-x)$.

The the increments of L\'evy $\alpha$-stable motion $dL_{t}^{\alpha}$
have the scaling property $dL_{at}^{\alpha}=a^{1/\alpha}dL_{t}^{\alpha}$
\cite{Janicki1994}. Changing the variable $x$ in Eq.~(\ref{eq:sde-2})
to the scaled variable $x_{s}=ax$ or introducing the scaled time
$t_{s}=a^{\alpha(\eta-1)}t$ one gets the same resulting equation.
Thus change of the scale of the variable $x$ and change of time scale
are equivalent, leading to the following scaling property of the transition
probability: 
\begin{equation}
aP(ax^{\prime},t|ax,0)=P(x^{\prime},a^{\mu}t|x,0)\,,\label{eq:scaling-1}
\end{equation}
with the exponent $\mu$ being 
\begin{equation}
\mu=\alpha(\eta-1)\,.
\end{equation}
As has been shown in Ref.~\cite{Ruseckas2014}, the power law steady
state PDF $P_{0}(x)\sim x^{-\lambda}$ and the scaling property of
the transition probability (\ref{eq:scaling-1}) lead to the power
law form PSD $S(f)\sim f^{-\beta}$ in a wide range of frequencies.
From the equation
\begin{equation}
\beta=1+(\lambda-3)/\mu\,,
\end{equation}
obtained in Ref.~\cite{Ruseckas2014}, it follows that the power-law
exponent in the PSD of the signal generated by SDE with L\'evy stable
noise (\ref{eq:sde-2}) is
\begin{equation}
\beta=1+\frac{\lambda-3}{\alpha(\eta-1)}\,.\label{eq:beta}
\end{equation}
This expression is the generalization of the expression for the power-law
exponent in the PSD with $\alpha=2$, obtained in Ref.~\cite{Kaulakys2006}.
As Eq.~(\ref{eq:beta}) shows, we get $1/f$ PSD when $\lambda=3$.

The presence of the restrictions at $x=x_{\mathrm{min}}$ and $x=x_{\mathrm{max}}$makes
the scaling (\ref{eq:scaling-1}) not exact and this limits the power
law part of the PSD to a finite range of frequencies $f_{\mathrm{min}}\ll f\ll f_{\mathrm{max}}$.
Similarly as in Ref.~\cite{Ruseckas2014} we can estimate the limiting
frequencies. Taking into account $x_{\mathrm{min}}$ and $x_{\mathrm{max}}$
the autocorrelation function has the scaling property \cite{Ruseckas2014}
\[
C(t;ax_{\mathrm{min}},ax_{\mathrm{max}})=a^{2}C(a^{\mu}t,x_{\mathrm{min}},x_{\mathrm{max}})\,.
\]
This equation means that time $t$ in the autocorrelation function
should enter only in combinations with the limiting values, $x_{\mathrm{min}}t^{1/\mu}$
and $x_{\mathrm{max}}t^{1/\mu}$. We can expect that the influence
of the limiting values can be neglected when the first combination
is small and the second large, that is when time $t$ is in the interval
$\sigma^{-\alpha}x_{\mathrm{max}}^{\alpha(1-\eta)}\ll t\ll\sigma^{-\alpha}x_{\mathrm{min}}^{\alpha(1-\eta)}$.
Then the frequency range where the PSD has $1/f^{\beta}$ behavior
can be estimated as 
\begin{equation}
\sigma^{\alpha}x_{\mathrm{min}}^{\alpha(\eta-1)}\ll2\pi f\ll\sigma^{\alpha}x_{\mathrm{max}}^{\alpha(\eta-1)}\,.\label{eq:freq-range}
\end{equation}
This equation shows that the frequency range grows with increasing
of the exponent $\eta$, the frequency range becomes zero when $\eta=1$.
By increasing the ratio $x_{\mathrm{max}}/x_{\mathrm{min}}$ one can
get arbitrarily wide range of the frequencies where the PSD has $1/f^{\beta}$
behavior. Note, that pure $1/f^{\beta}$ PSD is physically impossible
because the total power would be infinite. Therefore, we consider
signals with PSD having $1/f^{\beta}$ behavior only in some wide
intermediate region of frequencies, $f_{\mathrm{min}}\ll f\ll f_{\mathrm{max}}$,
whereas for small frequencies $f\ll f_{\mathrm{min}}$ PSD is bounded.

The power spectral density of the form $1/f^{\beta}$ is determined
mainly by power law behavior of the coefficients of SDE (\ref{eq:sde-2})
at large values of $x\gg x_{\mathrm{min}}$. Changing the coefficients
at small $x$, the spectrum preserves the power law behavior. The
modifications of the SDE (\ref{eq:sde-3}), (\ref{eq:sde-3a}) and
the introduction of negative values of the stochastic variable $x$
should not destroy the frequency region with $1/f^{\beta}$ behavior
of the power spectral density. This is confirmed by numerical solution
of the equations.

\section{Numerical examples}

\label{sec:numerical}When $\lambda=3$, we get that $\beta=1$ and
SDEs (\ref{eq:sde-power-1}), (\ref{eq:sde-3}), (\ref{eq:sde-3a})
should give a signal exhibiting $1/f$ noise. We will solve numerically
two cases, corresponding to Eqs.~(\ref{eq:sde-power-1}) and (\ref{eq:sde-3}),
with the index of stability of L\'evy stable noise $\alpha=1$ and
the power law exponent of the steady state PDF $\lambda=3$. Note,
that for this value of $\alpha$ the L\'evy $\alpha$-stable distribution
is the same as the Cauchy distribution. For simplicity we choose the
exponent in the noise amplitude $\eta$ such that the coefficient
$\gamma$, given by Eqs.~(\ref{eq:gamma-pos}) or (\ref{eq:gamma-pos-neg}),
becomes equal to $-1$. For the numerical solution we use Euler's
approximation, transforming differential equations to difference equations.
Eq.~(\ref{eq:sde-2}) leads to the following difference equation
\begin{equation}
x_{k+1}=x_{k}+\sigma^{\alpha}\gamma x_{k}^{\alpha(\eta-1)+1}h_{k}+x_{k}^{\eta}h_{k}^{1/\alpha}\xi_{k}^{\alpha}\,,\label{eq:discr}
\end{equation}
where $h_{k}=t_{k+1}-t_{k}$ is the time step and $\xi_{k}^{\alpha}$
is a random variable having $\alpha$-stable L\'evy distribution
with the characteristic function (\ref{eq:charact}). We can solve
Eq.~(\ref{eq:discr}) numerically with the constant step $h_{k}=\mathrm{const}$.
However, more effective method of solution of Eq.~(\ref{eq:discr})
is when the change of the variable $x_{k}$ in one step is proportional
to the value of the variable, as has been done solving SDE with Gaussian
noise in Ref.~\cite{Kaulakys2004}. Variable step of integration
\begin{equation}
h_{k}=\frac{\kappa^{\alpha}}{\sigma^{\alpha}}x_{k}^{-\alpha(\eta-1)}
\end{equation}
results in the equation
\begin{equation}
x_{k+1}=x_{k}+\kappa^{\alpha}\gamma x_{k}+\frac{\kappa}{\sigma}x_{k}\xi_{k}^{\alpha}\,.
\end{equation}
Here $\kappa\ll1$ is a small parameter. We include the reflective
boundaries at $x=x_{\mathrm{min}}$ and $x=x_{\mathrm{max}}$ using
the projection method \cite{Liu1995,Pettersson1995}. According to
the projection method, if the variable $x_{k+1}$ acquires the value
outside of the interval $[x_{\mathrm{min}},x_{\mathrm{max}}]$ then
the value of the nearest reflective boundary is assigned to $x_{k+1}$.

\begin{figure}
\includegraphics[width=0.33\textwidth]{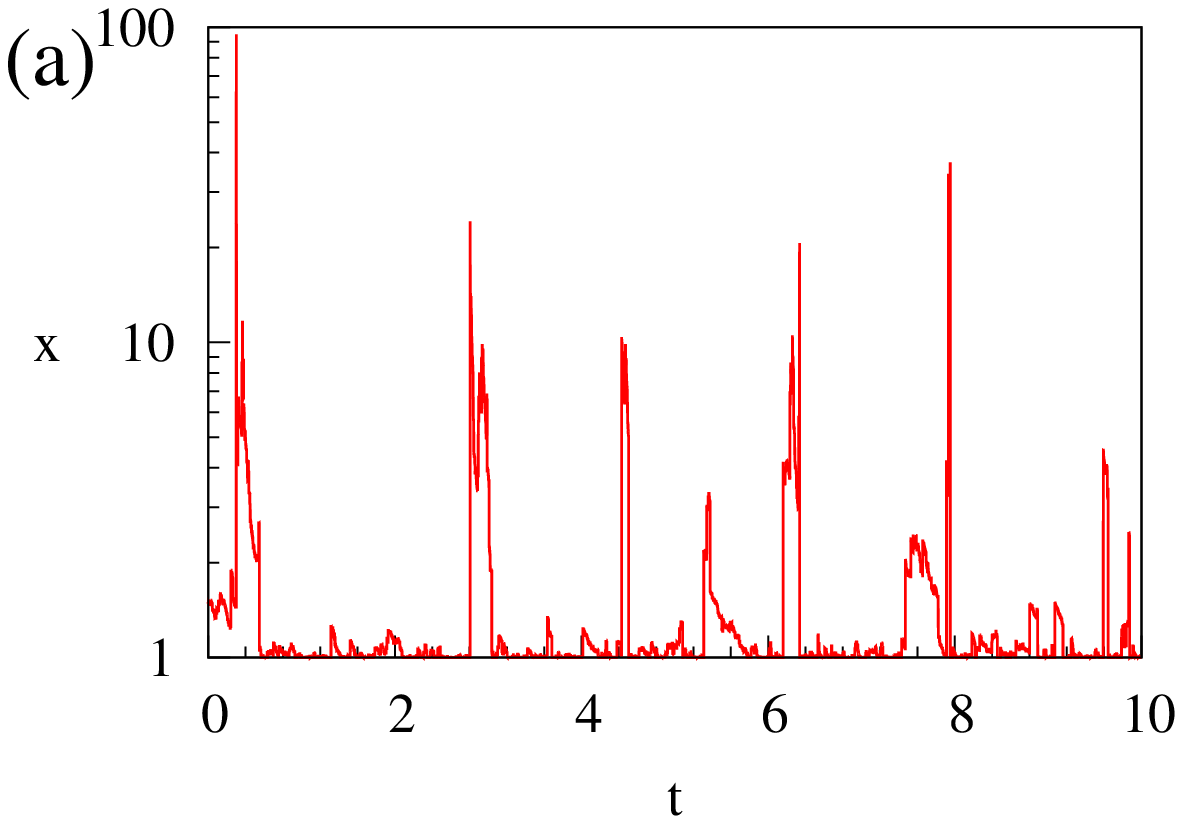}\includegraphics[width=0.33\textwidth]{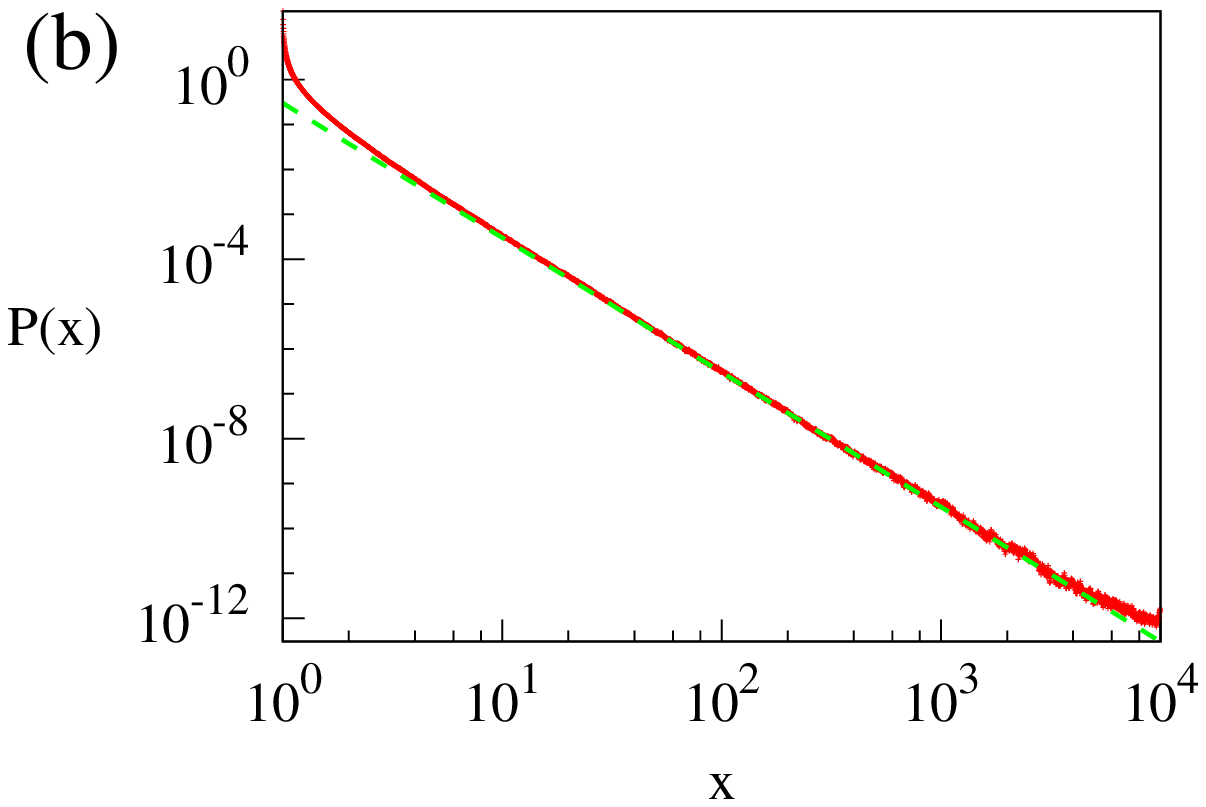}\includegraphics[width=0.33\textwidth]{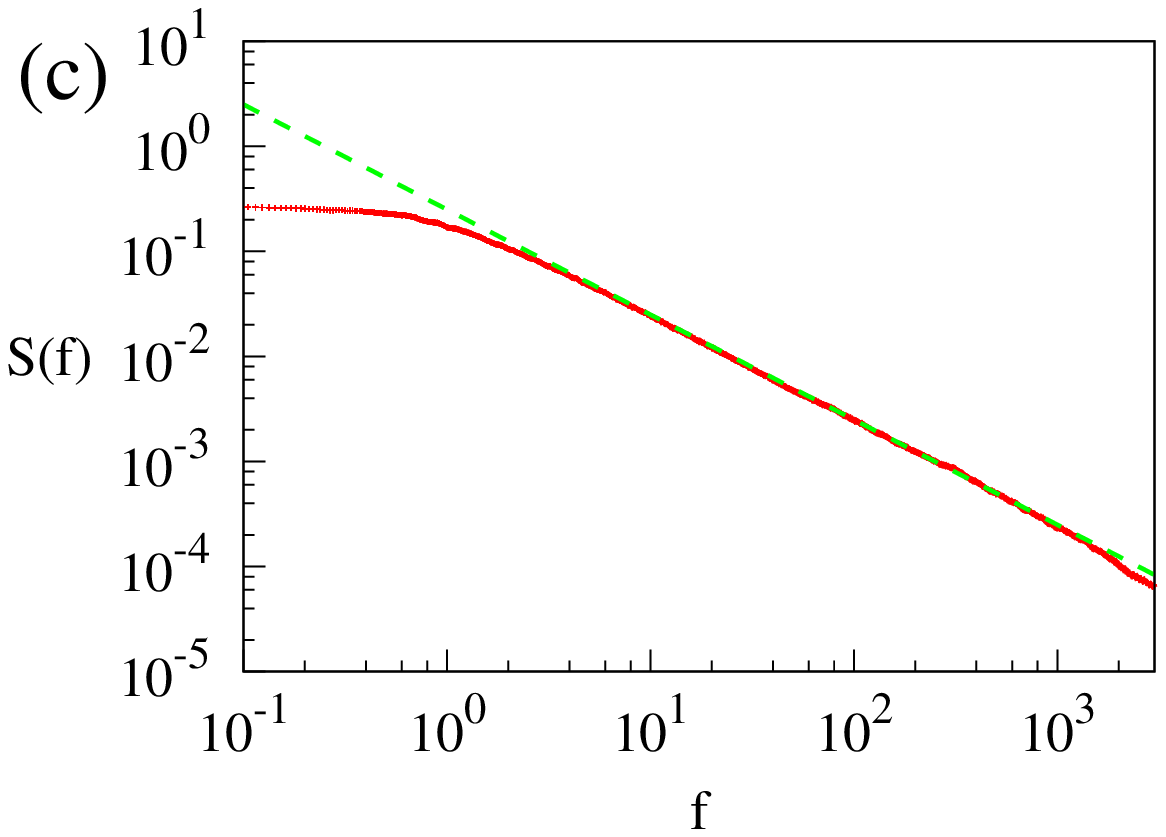}

\caption{(Color online) (a) Signal generated by SDE with L\'evy stable noise
(\ref{eq:numer-1}) with reflective boundaries at $x=x_{\mathrm{min}}$
and $x=x_{\mathrm{max}}$. (b) Steady state PDF $P_{0}(x)$ of the
signal. The dashed green line shows the slope $x^{-3}$. (c) Power
spectral density $S(f)$ of the signal. The dashed green line shows
the slope $1/f$. Parameters used are $x_{\mathrm{min}}=1$, $x_{\mathrm{max}}=10^{4}$,
$\sigma=1$.}
\label{fig:sol1}
\end{figure}

When $\alpha=1$, $\lambda=3$ and $\eta=9/4$, the SDE (\ref{eq:sde-power-1})
is
\begin{equation}
dx=-\sigma x^{9/4}dt+\sigma x^{9/4}dL_{t}^{1}\,.\label{eq:numer-1}
\end{equation}
The results obtained numerically solving this equation with reflective
boundaries at $x=x_{\mathrm{min}}$ and $x=x_{\mathrm{max}}$ are
shown in Fig.~\ref{fig:sol1}. A sample of the generated signal is
shown in Fig.~\ref{fig:sol1}a. The signal exhibits peaks or bursts,
corresponding to the large deviations of the variable $x$. Comparison
of the steady state PDF $P_{0}(x)$ and the PSD $S(f)$ with the analytical
estimations is presented in Fig.~\ref{fig:sol1}b and Fig.~\ref{fig:sol1}c.
There is quite good agreement of the numerical results with the analytical
expressions. In Fig.~\ref{fig:sol1}b we see that near the reflecting
boundaries the steady state PDF deviates from the power law prediction.
This increase of the steady state PDF near boundaries is typical for
equations with L\'evy stable noise having $\alpha<2$ \cite{Denisov2008a}.
The behavior of the steady state PDF near the reflecting boundaries
is similar to the behavior of the analytical expression obtained in
Ref.~\cite{Denisov2008a} for the simplest stochastic differential
equation L\'evy stable noise having constant noise amplitude and
zero drift.

A numerical solution of the equations confirms the presence of the
frequency region for which the PSD has $1/f$ dependence. The width
of this region can be increased by increasing the ration between the
minimum and the maximum values of the stochastic variable $x$. In
addition, the region in the PSD with the power law behavior depends
on $\alpha$ and the exponent $\eta$: the width increases with increasing
the difference $\eta-1$ and increasing $\alpha$; when $\eta=1$
then this width is zero. Such behavior is correctly predicted by Eq.~(\ref{eq:freq-range}).

\begin{figure}
\includegraphics[width=0.33\textwidth]{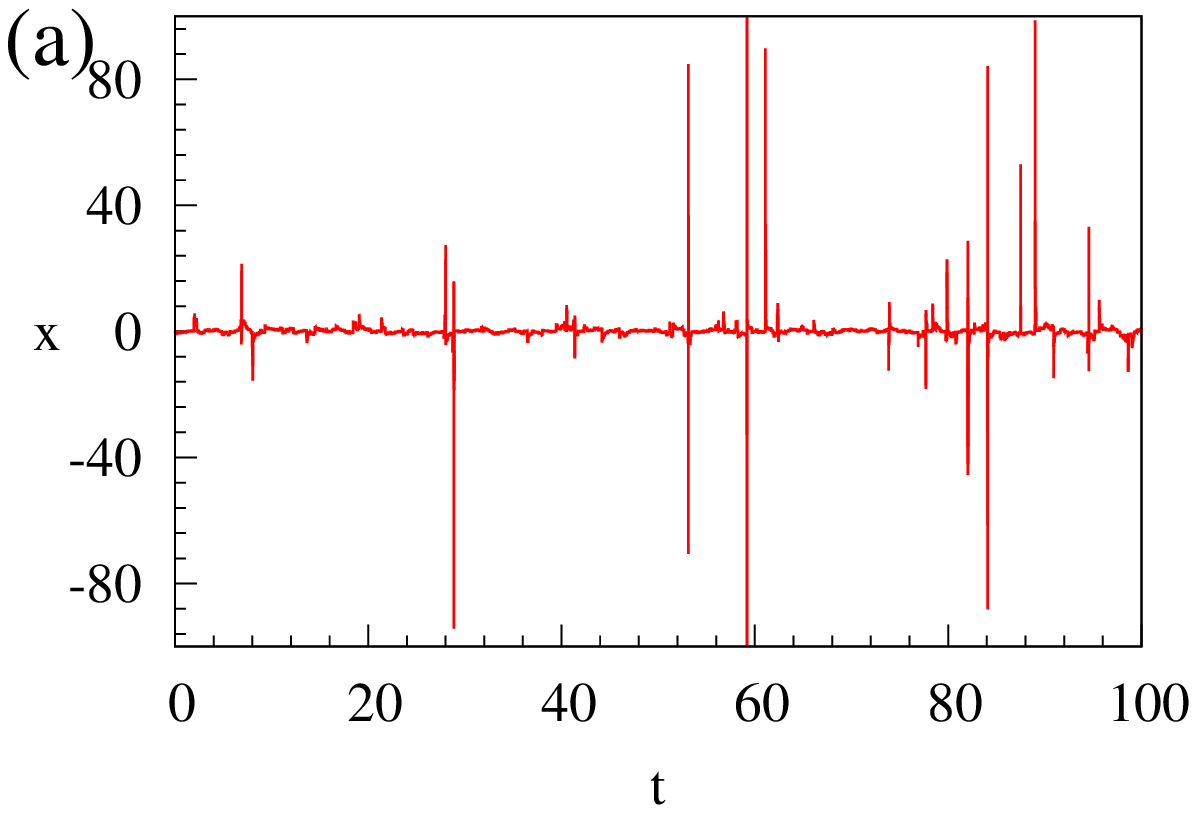}\includegraphics[width=0.33\textwidth]{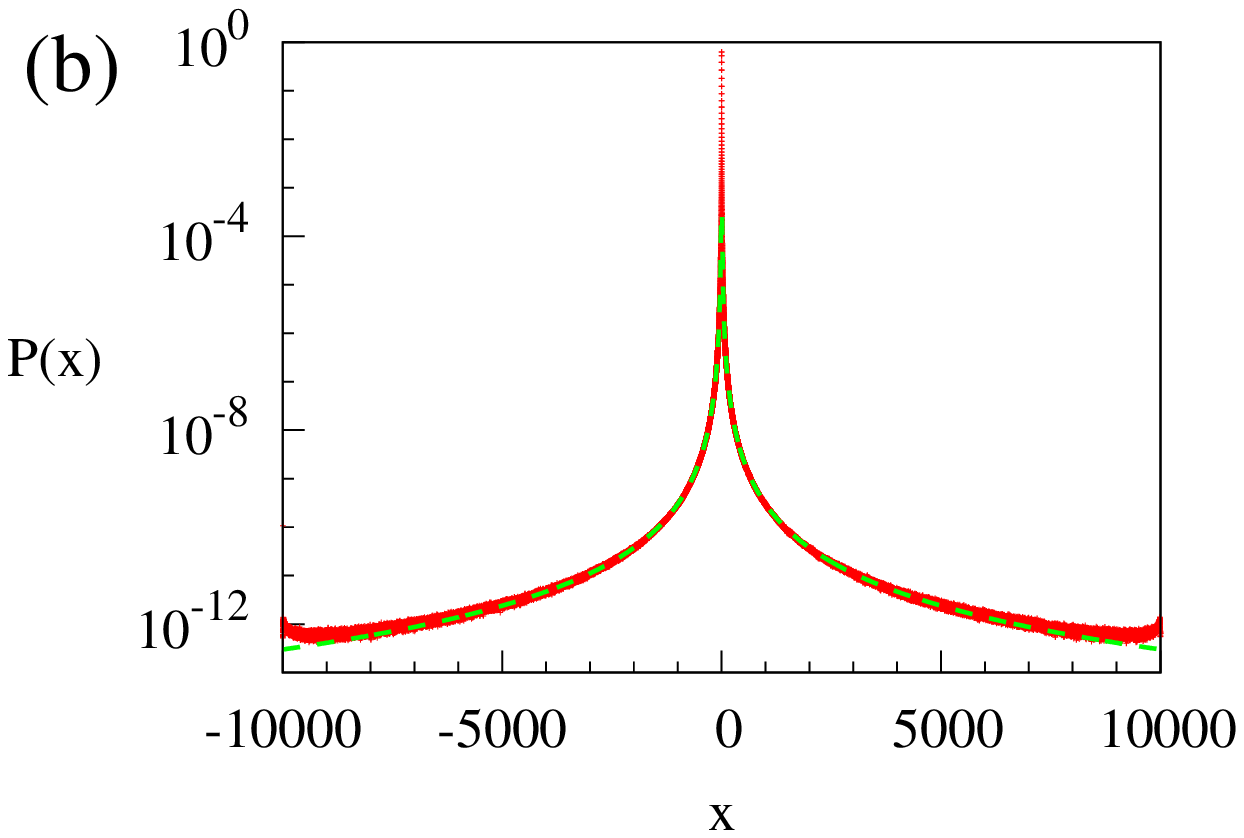}\includegraphics[width=0.33\textwidth]{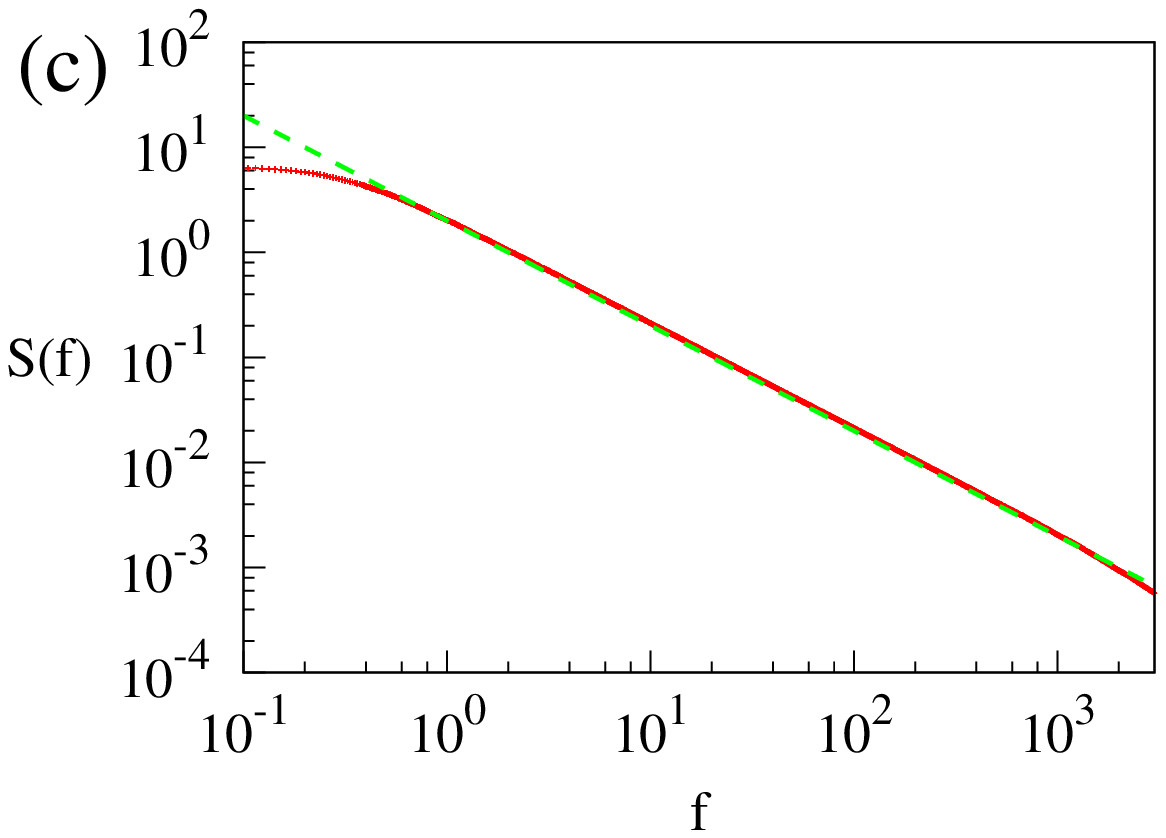}\caption{(Color online) (a) Signal generated by SDE with L\'evy stable noise
(\ref{eq:numer-2}). (b) Steady state PDF $P_{0}(x)$ of the signal.
The dashed green line shows the dependence on $x$ proportional to
$|x|^{-3}$. (c) Power spectral density $S(f)$ of the signal. The
dashed green line shows the slope $1/f$. Parameters used are $x_{0}=1$,
$x_{\mathrm{max}}=10^{4}$, $\sigma=1$.}
\label{fig:sol2}
\end{figure}

Similar schemes of numerical solution we use also for SDEs (\ref{eq:sde-3})
and (\ref{eq:sde-3a}). Euler's approximation with variable step of
integration
\begin{equation}
h_{k}=\frac{\kappa^{\alpha}}{\sigma^{\alpha}}(x_{0}^{2}+x_{k}^{2})^{-\frac{\alpha}{2}(\eta-1)}
\end{equation}
transforms SDE (\ref{eq:sde-3}) to the difference equation
\begin{equation}
x_{k+1}=x_{k}+\kappa^{\alpha}\gamma x_{k}+\frac{\kappa}{\sigma}\sqrt{x_{0}^{2}+x_{k}^{2}}\xi_{k}^{\alpha}\,.
\end{equation}
For SDE (\ref{eq:sde-3a}) we use the variable step of integration
\begin{equation}
h_{k}=\frac{\kappa^{\alpha}}{\sigma^{\alpha}}(x_{0}^{\alpha}+|x_{k}|^{\alpha})^{-(\eta-1)}
\end{equation}
resulting in the difference equation
\begin{equation}
x_{k+1}=x_{k}+\kappa^{\alpha}\gamma x_{k}+\frac{\kappa}{\sigma}(x_{0}^{\alpha}+|x_{k}|^{\alpha})^{\frac{1}{\alpha}}\xi_{k}^{\alpha}\,.
\end{equation}
Here $\kappa\ll1$ is a small parameter. Reflective boundaries at
$x=\pm x_{\mathrm{max}}$ we include using the projection method.

When $\alpha=1$ , $\lambda=3$ and $\eta=5/2$, the SDE (\ref{eq:sde-3})
with the coefficient $\gamma$ given by Eq.~(\ref{eq:gamma-pos-neg})
is
\begin{equation}
dx=-\sigma(x_{0}^{2}+x^{2})^{3/4}xdt+(x_{0}^{2}+x^{2})^{5/4}dL_{t}^{1}\,.\label{eq:numer-2}
\end{equation}
The results obtained numerically solving this equation with reflective
boundaries at $x=\pm x_{\mathrm{max}}$ are shown in Fig.~\ref{fig:sol2}.
A sample of the generated signal is shown in Fig.~\ref{fig:sol2}a.
Comparison of the steady state PDF $P_{0}(x)$ and the PSD $S(f)$
with the analytical estimations is presented in Fig.~\ref{fig:sol2}b
and Fig.~\ref{fig:sol2}c. There is quite good agreement of the numerical
results with the analytical expressions. As in the case with only
positive values of $x$, we see in Fig.~\ref{fig:sol1}b we see the
increase of the steady state PDF near the reflecting boundaries $x=\pm x_{\mathrm{max}}$
in comparison to the power law prediction. Numerical solution of Eq.~(\ref{eq:numer-2})
confirms the presence of the frequency region where the PSD has $1/f^{\beta}$
dependence.

\section{Discussion}

\label{sec:concl} L\'evy flights have been modeled using Langevin
equation with various subharmonic potentials and additive L\'evy
stable noise \cite{Jespersen1999,Brockmann2002,Chechkin2002,Brockmann2003lf}.
Proposed SDE~(\ref{eq:sde-power-1}) contains multiplicative L\'evy
stable noise and is a generalization of previous attempts to model
L\'evy flights. This SDE can be used to investigate L\'evy flights
in non-equilibrium and non-homogeneous environments, like porous media
and some cases of polymer chains \cite{Brockmann2003,Lomholt2005}.
If specific conditions given by Eq.~(\ref{eq:cond_pow_law-1}) are
satisfied, our model generates L\'evy flights exhibiting $1/f$ noise.
The drift term $a(x)$ in Eq.~(\ref{eq:sde-power-1}) represents
a subharmonic external force effecting the particle. L\'evy flights
in subharmonic potentials lead to various interesting phenomena such
as stochastic resonance in singe well potential \cite{Dybiec2009}.
The power law dependence of the diffusion coefficient $b^{2}(x)$
on the stochastic variable $x$ can be traced to the existence of
the energy flux due to temperature gradient in a bath. Long jumps
leading to L\'evy stable noise can arise from a complex scale free
structure of the bath as is in the case of enzyme diffusion on a polymer
\cite{Brockmann2003}. There are suggestions that the non-homogeneity
of the bath can be described by the dependence of the diffusion coefficient
on the particle coordinate $x$ \cite{Srokowski2009} and L\'evy
stable noise arises from the bath not being in an equilibrium.

In the case of Gaussian noise ($\alpha=2$) nonlinear SDE (\ref{eq:sde-power-1})
that generates signal with $1/f$ spectrum can be obtained from various
models. One of those models is a signal consisting form a sequence
of pulses with a Brownian motion of the inter-pulse durations \cite{Kaulakys2004,Kaulakys2006}.
This suggests that our more general form of the SDE could be obtained
from some kind of L\'evy motion of the inter-pulse durations. However,
we were unable to show this due to the complexity of It\^o formula
in case of equations driven by L\'evy process \cite{Jacobs2009}.
The special case of Eq.~(\ref{eq:sde-power-1}) for free particle
($a(x)=0$) with L\'evy stable noise having $\alpha<2$ has been
derived from coupled continuous time random walk (CTRW) models \cite{Srokowski2006},
when jumping rate $\nu$ of CTRW process depends on signal intensity
as $\nu(x)=x^{\alpha\eta}$, $x>0$. However, such derivation is quite
complex and does little to help the understanding what kind of physical
phenomena can be approximated by multiplicative L\'evy stable noise.
Thus instead of searching for underlying models in this article we
have chosen an simpler approach: we have derived nonlinear SDEs using
a simple reasoning about scaling properties of the steady state PDF.

Taking into account of the scaling properties of the signal is one
of the advantages of our model. In many theoretical models, such as
diffusion of the particle in a fractal turbulence \cite{Takayasu1984},
ecological population density with fluctuating volume of resources
\cite{Alexander2012}, dynamics of two competing species \cite{LaCognata2010}
and tumor growth \cite{jiang2012}, an existence of L\'evy stable
noise instead of Gaussian noise is simply assumed. Such assumption
might be incorrect, because the change of statistical properties of
the noise change the scaling properties of the signal. In order to
preserve original scaling properties of the signal the drift $a(x)$
or diffusion $b^{2}(x)$ coefficients must be changed as well. The
required drift coefficient $a(x)$ can be found similarly as in Section~\ref{sec:power}.
The scaling properties can be extracted from time series using fluctuation
analysis methods \cite{Weron2005}.

In summary, we have proposed nonlinear SDEs with L\'evy stable noise
and generating signals exhibiting $1/f$ noise in any desirably wide
range of frequency. Proposed SDEs (\ref{eq:sde-power-1}), (\ref{eq:sde-3})
and (\ref{eq:sde-3a}) are a generalization of nonlinear SDEs driven
by Gaussian noise and generating signals with $1/f$ PSD. The generalized
equations can be obtained by replacing the Gaussian noise with the
L\'evy stable noise and changing the drift term to preserve statistical
properties of the generated signal. We have investigated two cases:
in the first case the stochastic variable can acquire only positive
values (SDE (\ref{eq:sde-power-1})), in the second case the stochastic
variable can also be negative (SDEs (\ref{eq:sde-3}) and (\ref{eq:sde-3a})).
In contrast to the SDEs with the Gaussian noise, the constant in the
drift term, given by Eqs.~(\ref{eq:gamma-pos}) and (\ref{eq:gamma-pos-neg}),
is different in those two cases and becomes the same only for $\alpha=2$.


%

\end{document}